\documentclass[a4paper,onecolumn,unpublished,allowtoday]{quantumarticle}
\pdfoutput=1
\usepackage{standalone}
\usepackage[page,header]{appendix}
\usepackage{titletoc}
\usepackage{dsfont}
\usepackage[utf8]{inputenc}
\usepackage[english]{babel}
\usepackage[T1]{fontenc}
\usepackage{float}
\usepackage[10pt]{moresize}
\usepackage{amsmath}
\usepackage{amssymb}
\usepackage{amsthm}
\usepackage{physics}
\usepackage{wrapfig}
\usepackage{mathtools}
\usepackage{hyperref}
\usepackage{pgfplots}
\usepackage{delarray}
\pgfplotsset{width=10cm,compat=1.9}
\usepgfplotslibrary{groupplots, fillbetween}
\usepackage{cleveref}

\crefname{equation}{Equation}{Equation}
\Crefname{equation}{Equation}{Equations}

\hypersetup{
    colorlinks,
    linkcolor=myblue,
    citecolor=myblue,
    filecolor=myblue,
    urlcolor=myblue,
    pdftitle={Coherent attacks are stronger than collective attacks on DIQKD with random postselection},
    pdfauthor={Martin Sandfuchs and Ramona Wolf}
}

\definecolor{myred}{RGB}{236, 17, 0}
\definecolor{myblue}{RGB}{10, 88, 153}
\definecolor{mygreen}{RGB}{26, 152, 81}
\definecolor{myorange}{RGB}{236, 137, 0}
\definecolor{LightGray}{RGB}{220,220,220}
\definecolor{LinkColor}{RGB}{167,20,49}

\crefname{figure}{Figure}{Figures}
\Crefname{figure}{Figure}{Figures}

\title{Coherent attacks are stronger than collective attacks on DIQKD with random postselection}
\author{Martin Sandfuchs}
\email{martisan@phys.ethz.ch}
\author{Ramona Wolf}
\email{rawolf@phys.ethz.ch}
\affiliation{Institute for Theoretical Physics, ETH Zürich, Wolfgang-Pauli-Str.\ 27, 8093 Zürich, Switzerland}
\begin{document}
\maketitle

\fontfamily{lmr}\selectfont

\begin{abstract}
	In a recent paper \cite{Liu_2022}, the authors report on the implementation of a device-independent QKD protocol with random postselection, which was originally proposed in \cite{Xu_2022}. Both works only provide a security proof against collective attacks, leaving open the question whether the protocol is secure against coherent attacks. In this note, we report on an attack on this protocol that demonstrates that coherent attacks are, in fact, stronger than collective attacks. 
\end{abstract}

\section{Introduction}

Device-independent quantum key distribution (DIQKD) poses significant challenges on experimental realisations, since it requires the execution of a loophole-free Bell test. Recently, three different research groups have reported on the first successful implementations of DIQKD \cite{Nadlinger2022,Zhang2022,Liu_2022}. Notably, one of these implementations \cite{Liu_2022} employs a protocol that includes a random postselection step, which was originally proposed in \cite{Xu_2022}. This entails that with a certain probability, a protocol round that has produced an output of $1$ is discarded and publicly announced. However, the security of this protocol has, to date, only been proven against collective attacks in the asymptotic regime, leaving a crucial question unanswered: Is the protocol secure against coherent attacks? 

In this note, we present a coherent attack (also called general attack) against the protocol that is stronger than all collective attacks. This has two significant consequences. Firstly, it implies that all techniques used to prove security against general attacks via a reduction to collective attacks cannot be applied to this protocol. This concerns standard techniques such as the entropy accumulation theorem \cite{Dupuis2020,Metger2022} and the quantum de Finetti theorem \cite{Renner2007,Renner2008}. Secondly, our attack sets an upper bound on the actual certifiable key rate in the case of general attacks, which means that the previously predicted certifiable secure key rate based on prior analyses is overestimated.

\section{Description of the random postselection protocol}

The random postselection (RPS) protocol \cite{Xu_2022} works in a similar way to the usual DIQKD protocols based on the CHSH inequality \cite{Pironio2009}: Alice and Bob are each in possession of a device with two and three inputs, respectively. The devices are fed with inputs $x \in \{1, 2\}$ (for Alice) and $y \in \{1, 2, 3\}$ (for Bob) and produce outputs $a, b \in \{0, 1\}$. The input-output distribution is described by a conditional probability distribution $p(ab|xy)$. In a random subset of the rounds, Alice and Bob collect statistics to test the devices for honesty. In contrast to CHSH-based DIQKD protocols, however, they collect bounds on the full distribution $p(ab|xy)$ rather than just the CHSH value. The other rounds constitute the key rounds, where Alice chooses $x=0$ and Bob chooses $y=3$.

Any realistic implementation of a DIQKD protocol will have to deal with noise, in particular losses of the qubits. In the RPS protocol, lost signals are assigned the outcomes $a = 1$ and $b = 1$ (i.e., binning to $1$). The RPS protocol then post-processes the measurement outcomes in the key rounds (i.e., $x = 1$, $y = 3$) by Alice and Bob randomly keeping (discarding) outputs $a=1$ or $b=1$ with some probability $p$ ($1 - p$). The outcomes $a=0$ and $b=0$ are never discarded. A round is discarded if either party decides to discard their output. This requires Alice and Bob to communicate the discarding and non-discarding information publicly, i.e., this information will become known to Eve.

The security of this protocol against collective attacks has been established in \cite{Xu_2022,Liu_2022}. However, it remains unclear whether this protocol is secure against general attacks. The biggest obstacle in answering this question is that standard proof techniques cannot be applied due to the structure of the protocol: In case of the entropy accumulation theorem \cite{Dupuis2020}, the protocol violates the Markov condition on the quantum channels (or, in the generalised version of the theorem \cite{Metger2022}, it violates the no-signalling condition) because the public announcement whether a round is discarded contains information on the outcome of that round. The quantum de Finetti theorem \cite{Renner2007,Renner2008} generally cannot be applied in the device-independent setting. In fact, no method that works via a reduction from coherent to collective attacks can be applied to this protocol, which we demonstrate by constructing a coherent attack that is stronger than all collective attacks.

\section{Description of the attack}
\label{sec:attack}

In this section, we describe an explicit attack on the protocol. The idea is to modify the behaviour of the devices in order to reveal information on the key to the eavesdropper. In particular, the attack exploits that the public information whether a round is discarded discloses information on the outcome of that round.

Throughout this paper we denote by $p_0 \coloneqq p(a=0|x=1)$ the probability that Alice's device produces the output $a=0$ in a key round ($x=1$). Similarly, we write $p_1 \coloneqq p(a=1|x=1) = 1 - p_0$. In the protocol, Alice and Bob's devices each receive half of an entangled Bell state $\ket{\psi}_{AB}$. Alice's device always performs the usual CHSH measurements:
\begin{align}
    M_1 = Z, \quad M_2 = X,
\end{align}
where $Z$ and $X$ are Pauli operators and the subscripts denote the choice of measurement for any given input $x \in \{1, 2\}$. 
If Bob's device receives as input $y \in \{1, 2\}$ it also performs the usual CHSH measurements:
\begin{align}
    N_1 = \frac{Z + X}{\sqrt{2}}, \quad N_2 = \frac{Z - X}{\sqrt{2}}.
\end{align}

If $y=3$, the round is a key round. Bob's device holds a counter $C$ which counts the number of times it received the input $y=3$. Given the input $y=3$, Bob's device reads the value of $C$ and acts according to the following strategy: If $C$ is even, it measures the operator $N_3 = Z$ (note that for the first round we have $C=0$). If $C$ is odd, the device outputs the result from the previous round where $y=3$, which the device keeps in its memory. In summary, Bob's device acts honestly in even (key)-rounds and simply repeats the output of the previous (key)-round during odd rounds.

Since the devices still maximally violate the CHSH inequality, we know that Alice and Bob must share a Bell pair. This means that the attack is purely classical since it only exploits the classical information leaked by Alice and Bob to learn something about the previous rounds. It is also worth noting that in rounds where $C$ is even (i.e., the device behaves honestly) Alice and Bob's outputs are perfectly correlated. We will use this later to simplify some computations.

\section{Analysis of the attack}

We now analyse the attack presented above and show that it is, in fact, stronger than all collective attacks. To compare different classes of attacks, one has to compare the respective conditional von Neumann entropy $H(A|E\Sigma)$, which describes the knowledge of the attacker (which consists of a quantum system $E$ and the classical communication $\Sigma$) by on the generated key $A$ in the asymptotic limit.

\subsection{Bound for collective attacks}

To compute a bound on the entropy for the collective attacks (also called \emph{iid attacks}, which is short for independent and identically distributed), we follow the procedure presented in \cite{Liu_2022} by evaluating the von Neumann entropy using the method developed in \cite{Brown_2021}. With this we can compute a lower bound on $H_{\mathrm{iid}}(A|E \Sigma)$, where $\Sigma$ denotes the public classical communication exchanged during the protocol (which includes the information whether a round has been discarded).

\subsection{Analysis of the coherent attack}

For the analysis of the coherent attack we need to consider pairs of key rounds. In the first round the device behaves honestly while in the second round Bob's device repeats the output from the first round. For this we denote by $A_1 \in \{0, 1, \bot\}$ the random variable describing Alice's output in the first round (after Alice post selects her bit) and by $A_2 \in \{0, 1, \bot\}$ Alice's output in the second round (again, after Alice has performed the post selection on her output). Similarly, we write $S_1, T_1, S_2, T_2 \in \{\bot, \top\}$ for the abort and non-abort side-information leaked during the two rounds: $S_i$ ($T_i$) describes the information whether Alice (Bob) discards the outcome of round $i \in \{1, 2\}$. We denote by $p(a_2, s_2, t_2, a_1, s_1, t_1)$ the joint probability distribution of obtaining a given outcome during two rounds of the coherent attack. Note that if a round $i$ is discarded, the output of that round is set to $a_i=\perp$. We can write
\begin{equation}
\begin{aligned}
    p(a_2, s_2, t_2, a_1, s_1, t_1) &= p(a_2, s_2, t_2|a_1, s_1, t_1) \cdot p(a_1, s_1, t_1) \\
    &= p(a_2, s_2, t_2|a_1) \cdot p(a_1, s_1, t_1),
\end{aligned}
\end{equation}
where we have noted that neither device has access to $s_1$ or $t_1$ (since the decision which rounds are discarded is only made after all measurements have been completed). Next, we need to compute the distribution $p(a_1, s_1, t_1)$ of the first round. Many of the components are zero (for instance if $s_1=\bot$ then we know for sure that $a_1=\bot$). Therefore, we only write down the non-zero components here:
\begin{equation}
\begin{aligned}
    p(a_1=0, s_1=\top, t_1=\top) &= p_0, \\
    p(a_1=1, s_1=\top, t_1=\top) &= p_1\cdot p^2, \\
    p(a_1=\bot, s_1=\top, t_1=\bot) &= p_1 \cdot p\cdot (1 - p), \\
    p(a_1=\bot, s_1=\bot, t_1=\top) &= p_1 \cdot(1 - p)\cdot p, \\
    p(a_1=\bot, s_1=\bot, t_1=\bot) &= p_1 \cdot(1 - p)^2.
\end{aligned}
\end{equation}

\begin{figure}[t]
	\centering
	\begin{tikzpicture}
		\begin{axis}[
			width = 0.7\linewidth,
			xlabel={$p$},
			ylabel={entropy},
			xmin=0, xmax=1,
			ymin=0, ymax=1,
			ymajorgrids=true,
			grid style=dashed,
			legend pos=north west,
			every mark=none,
			legend cell align={left},
			scale=1.0,
			]
			\addplot+[solid, myred, very thick, mark=none] table[x=p, y=ent, col sep=comma]{entropies_iid.csv};
			\addlegendentry{$H_{\mathrm{iid}}(A|E \Sigma)$}
			
			\addplot+[solid, myblue, very thick, mark=none] table[x=p, y=ent, col sep=comma]{entropies_attack.csv};
			\addlegendentry{$H_{\mathrm{attack}}(A|E \Sigma)$}
		\end{axis}
	\end{tikzpicture}
	
	\caption{\label{fig:comparison}Comparison of the entropies for collective attacks, $H_{\mathrm{iid}}(A|E\Sigma)$, and the coherent attack presented in Section~\ref{sec:attack}, $H_{\mathrm{attack}}(A|E\Sigma)$. The parameter $p$ is the probability that a key round with outcome $1$ is kept, and the plot is generated for $p_0\coloneqq p(a=0|x=1)=0.5$.}
\end{figure}
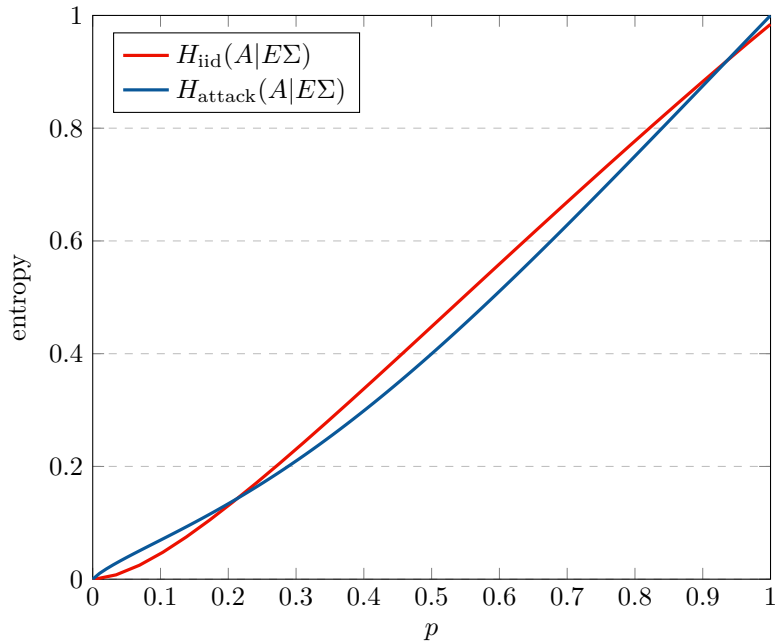

Next, we compute the distribution of the second round conditioned on receiving output $a_1$ in the first round. If $a_1 \in \{1, \bot\}$, then the output of the previous round before postselection must have been $1$ (for both Alice and Bob since they are perfectly correlated). This means that Bob's device outputs the value $1$. Therefore, we get (again only writing the non-zero components):
\begin{equation}
\begin{aligned}
    p(a_2=0, s_2=\top, t_2=\top|a_1 \in \{1, \bot\}) &= p_0\cdot p, \\
    p(a_2=1, s_2=\top, t_2=\top|a_1 \in \{1, \bot\}) &= p_1\cdot p^2, \\
    p(a_2=\bot, s_2=\bot, t_2=\top|a_1 \in \{1, \bot\}) &= p_1\cdot (1 - p) \cdot p, \\
    p(a_2=\bot, s_2=\top, t_2=\bot|a_1 \in \{1, \bot\}) &= (p_0 + p_1\cdot p)\cdot (1 - p), \\
    p(a_2=\bot, s_2=\bot, t_2=\bot|a_1 \in \{1, \bot\}) &= p_1 \cdot (1 - p)^2.
\end{aligned}
\end{equation}
On the other hand, if $a_1 = 0$ we get:
\begin{equation}
\begin{aligned}
    p(a_2=0, s_2=\top, t_2=\top|a_1=0) &= p_0, \\
    p(a_2=1, s_2=\top, t_2=\top|a_1=0) &= p_1 \cdot p, \\
    p(a_2=\bot, s_2=\bot, t_2=\top|a_1=0) &= p_1 \cdot (1 - p).
\end{aligned}
\end{equation}
With this distribution we can directly compute $H_\mathrm{attack}(A_1 A_2|S_1S_2T_1T_2)$, since it is a purely classical entropy.

\subsection{Comparison of collective and coherent attack}

The crucial question is now whether $H_\mathrm{iid}(A|E\Sigma) > H_\mathrm{attack}(A|E \Sigma)$ or, equivalently, whether $H_\mathrm{iid}(A|E\Sigma) > H_\mathrm{attack}(A_1 A_2|S_1S_2T_1T_2)/2$.
Of particular interest to us is the situation where $p_0 = 0.5$ because this is the situation for the states and operators stated in Section~\ref{sec:attack}. The result is shown in \cref{fig:comparison}.

As can be seen from the figure, the coherent attack we have presented here  outperforms all iid attacks (in some parameter range). This shows that the performance of the protocol against general attacks is, in general, worse than against the collective attacks studied so far. Thus, for the random postselection DIQKD protocol coherent attacks are, in fact, stronger than collective ones.

\section*{Acknowledgements}
We thank Mads Friis Frand-Madsen and Renato Renner for helpful discussions. This work was supported by the Air Force Office of Scientific Research (AFOSR), grant No.~FA9550-19-1-0202, the QuantERA project eDICT, the National Centre of Competence in Research SwissMAP, and the Quantum Center at ETH Zurich.

\bibliographystyle{halpha}
\bibliography{sources}

\end{document}